\documentclass[twocolumn,10pt]{article}
\usepackage[margin=1in]{geometry}
\usepackage{cite}
\usepackage{amsmath,amssymb,amsfonts}
\usepackage{algorithmic}
\usepackage{graphicx}
\usepackage{textcomp}
\usepackage{hyperref}
\usepackage{etoolbox}
\usepackage{caption}

\AtBeginEnvironment{table}{\footnotesize}
\AtBeginEnvironment{table*}{\footnotesize}
\def\BibTeX{{\rm B\kern-.05em{\sc i\kern-.025em b}\kern-.08em
    T\kern-.1667em\lower.7ex\hbox{E}\kern-.125emX}}
\begin{document}
\title{Task-Adaptive Low-Dose CT Reconstruction}
\author{
Necati Sefercioglu$^{1}$, Mehmet Ozan Unal$^{2}$, Metin Ertas, Isa Yildirim$^{2}$\\
\vspace{2mm}\\
\begin{minipage}[t]{0.48\textwidth}
\centering
{\small $^{1}$Department of Data Science and Analytics,}\\
{\small Istanbul Technical University, Istanbul, Turkiye}\\
{\small (sefercioglu21@itu.edu.tr)}
\end{minipage}
\hfill
\begin{minipage}[t]{0.48\textwidth}
\centering
{\small $^{2}$Department of Electronics and Communication Engineering,}\\
{\small Istanbul Technical University, Istanbul, Turkiye}\\
{\small (unalmehmet@itu.edu.tr, iyildirim@itu.edu.tr)}
\end{minipage}
}
\date{}

\maketitle

\renewcommand{\thefootnote}{}
\footnotetext{Computing resources used in this work were provided by the National Center for High Performance Computing of Turkey (UHeM) under grant number 4022252025.}
\renewcommand{\thefootnote}{\arabic{footnote}}

\begin{abstract}
Deep learning-based low-dose computed tomography reconstruction methods already achieve high performance 
on standard image quality metrics like peak signal-to-noise ratio and structural similarity index measure. 
Yet, they frequently fail to preserve the critical anatomical details needed for diagnostic
tasks. This fundamental limitation hinders their clinical applicability despite their high metric scores. 
We propose a novel task-adaptive reconstruction framework that addresses this gap by incorporating a frozen 
pre-trained task network as a regularization term in the reconstruction loss function. Unlike existing 
joint-training approaches that simultaneously optimize both reconstruction and task networks, and risk diverging 
from satisfactory reconstructions, our method leverages a pre-trained task model to guide reconstruction training 
while still maintaining diagnostic quality. We validate our framework on a liver and liver tumor segmentation 
task. Our task-adaptive models achieve Dice scores up to 0.707, approaching the performance of full-dose scans 
(0.874), and substantially outperforming joint-training approaches (0.331) and traditional reconstruction methods 
(0.626). Critically, our framework can be integrated into any existing deep learning-based reconstruction 
model through simple loss function modification, enabling widespread adoption for task-adaptive optimization in 
clinical practice. Our codes are available at: \url{https://github.com/itu-biai/task_adaptive_ct}
\end{abstract}


\section{Introduction}
\label{sec:introduction}
Low-dose computed tomography (LDCT) has emerged as a critical advancement in medical imaging, 
offering significant reduction in radiation exposure while maintaining diagnostic capabilities. Medical studies 
provide strong clinical evidence that LDCT screening leads to significant mortality reduction for some cancer 
types \cite{b1}. For instance, lung cancer screening in high-risk populations shows that early detection via LDCT reduces 
lung cancer mortality by about 20-24\% compared to chest X-rays or no screening \cite{b2}. Due to its implications 
for public health, there exists a strong motivation for LDCT, and ALARA principle has become the widespread 
standard for radiation exposure in medical imaging, standing for ``as low as reasonably achievable'' \cite{b3}.

There exist two main strategies to obtain low-dose scans. Radiation exposure can be reduced both by decreasing the 
scanning angles to acquire fewer projections, called sparse-view CT, and by reducing X-ray tube current 
intensity, called dose reduction \cite{b4}. Regardless of the approach, dose reduction introduces artifacts and noise
into the reconstructions, decreasing the diagnostic capabilities of the obtained scans \cite{b5}. Accordingly, low-dose 
reconstruction methods have emerged to mitigate these effects.

Initial successful work on LDCT reconstruction relied on conventional signal processing methods and iterative methods 
\cite{fbp}, \cite{art}, \cite{sart}, \cite{mbir}, \cite{asir}, \cite{tv}, \cite{bm3d}. 
When deep learning-based methods proved their success in classic image processing problems, the focus on LDCT 
research also shifted to deep learning-based reconstructions, which provided state-of-the-art results 
\cite{fbpconvnet}, \cite{redcnn}, \cite{wgan-vgg}, \cite{du-gan}, \cite{learned-primal-dual}, 
\cite{solving-ill-posed-inverse-problems}, \cite{deep-convolutional-framelet-denosing}. 

Most commonly used metrics in low-dose reconstruction literature are peak signal-to-noise ratio (PSNR) and 
structural similarity index measure (SSIM). PSNR metric measures the ratio between the maximum possible
signal power and the power of corrupting noise that affects the quality of the image. It is purely based 
on pixel-by-pixel differences and doesn't consider structural information. However, SSIM calculates the structural 
similarity between two images by comparing local patterns of pixel intensities, considering luminance, 
contrast, and structure \cite{ssim}, \cite{c1}. 

Deep learning-based reconstructions usually outperform traditional methods in both metrics, while also providing 
higher quality visuals with decreased noise and sharper details. Yet, it doesn't necessarily mean that 
these deep learning powered reconstructions perform well in clinical practice. PSNR is highly sensitive to pixel-wise 
differences and doesn't reflect the diagnostic quality, and SSIM focuses on structural similarity but may still miss subtle 
pathological changes. Accordingly, small lesions or pathological changes might be smoothed out while maintaining 
high PSNR and high SSIM, edge preservation of critical anatomical structures may be compromised despite good metric 
scores, and texture changes that are clinically significant might be altered while keeping the metrics high \cite{d1}, 
\cite{d2}, \cite{d3}, \cite{d4}.

Radiologists might perform various tasks after low-dose reconstruction. They can conduct detection tasks
like tumor detection and lesion detection, segmentation tasks like organ segmentation and tumor 
segmentation, and classification tasks like tumor classification, among various other tasks \cite{e1}. 
Currently, deep learning-based methods can also conduct these tasks with notable performances when fed full-dose scans 
as inputs \cite{e2}, \cite{e3}, \cite{e4}. 

Surprisingly, there remains a great gap in the literature about performances of LDCT reconstructions 
in the aforementioned tasks. Most of the previous work on LDCT reconstruction literature focus on increasing 
performance in PSNR and SSIM metrics, and abandon any further assessment and modification. However, as described 
above, these metrics are not sufficient to conclude that obtained reconstructions maintain enough 
information to conduct diagnostic tasks from. We acknowledge that PSNR and SSIM metrics are useful for evaluating 
low-dose reconstruction models, but emphasize that they are not sufficient to assess the potential clinical 
applicability of the models. In this work we won't present a new metric, but we advise the audience to research 
the recent literature about medical image quality assessment \cite{d1}, \cite{d2}, \cite{d3}, \cite{d4}. 
This research focuses on enhancing the clinical interpretability of the low-dose scans during the reconstruction 
process. In other words, we focus on creating reconstructions with enhanced performances in diagnostic tasks, 
what is usually called task-adapted, task-oriented, or task-informed reconstruction in the literature 
\cite{task-adapted-reconstruction}, \cite{task-oriented}, \cite{empirical-evidence}, \cite{task-informed}, 
\cite{impact-of-task-information}.

To the best of our knowledge, Adler et al. \cite{task-adapted-reconstruction} proposed the foundational work in 
task-adapted reconstruction literature, applying a joint-training framework that trains both reconstruction 
and task networks simultaneously with a corresponding hyperparameter controlling the weight between reconstruction 
and task losses. Valat et al. \cite{empirical-evidence} and Li et al. \cite{task-informed}, 
\cite{impact-of-task-information} employ this joint-training approach and provide further evidence that 
joint-training framework is working as intended using systematic evaluations. Zhang et al. \cite{task-oriented} 
propose a novel approach where they introduce a task-oriented loss in the WGAN \cite{wgan}. They add a task-oriented 
loss, coming from a frozen pre-trained task network and a reconstruction loss, into the discriminator loss of the 
WGAN. Weight hyperparameter is only present for the reconstruction loss, a tradeoff doesn't exist between 
reconstruction loss and task loss like the joint-training approach. 

Joint training approach of Adler et al. \cite{task-adapted-reconstruction} provides significant results about the 
potential of task-adapted approaches to reach reconstructions with enhanced performances in diagnostic tasks. 
They provide strong evidence that if an optimal task-adaptation is conducted, reconstructions can still contain 
relevant information about the task in hand, while still having satisfactory low-dose reconstructions. However, 
joint-training approach is too sensitive with the weight hyperparameter and the tradeoff is too strong. Moreover, 
one can completely diverge from satisfactory reconstructions trying to optimize the task performance. The main goal 
in task-adaptive reconstruction is to contain better information about the diagnostic task in reconstructed scans, 
it is not to conduct the task itself, so this remains a significant issue.

Task-oriented approach of Zhang et al. \cite{task-oriented} uses a pre-trained task network as a guide for 
training of the reconstruction network. This is an interesting and promising approach because the model doesn't have 
the issue of diverging from satisfactory reconstructions, while also still incorporating task knowledge. Results 
show that quality enhancement on region of interests---segmentation areas for this work---are more significant 
than on whole images, providing evidence that information about the task is actually maintained in reconstructions. However, 
there are still significant issues. First of all, this work is conducted specifically for WGAN 
\cite{wgan}, so generalization of the ideas about using pre-trained networks is not possible. Next, there is no 
hyperparameter controlling the balance between task loss and reconstruction loss, meaning that we can't control the 
focus of the model.

To address the limitations of current approaches, we propose a novel method, which we name task-adaptive 
reconstruction. We add a pre-trained task network as a regularizer in the loss function of reconstruction network, 
with an alpha hyperparameter controlling the weight between reconstruction loss and task loss. The pre-trained 
task network is frozen. It is only used to guide the optimization of reconstruction. Extensive experiments 
show that our proposed task-adaptive reconstruction method significantly increases task performances of 
the reconstructed LDCT scans. Following previous work \cite{task-adapted-reconstruction}, \cite{task-oriented}, 
\cite{empirical-evidence}, we focus on segmentation task to conduct our experiments.

Our main contributions are as follows:

\begin{itemize}

    \item We introduce a novel task-adaptive reconstruction framework that incorporates a pre-trained task network as a 
    regularization term in the loss function of the reconstruction network, with a corresponding alpha hyperparameter 
    controlling the weight between the reconstruction and associated task. 

    \item We propose an approach that can be easily integrated to any existing deep learning-based LDCT 
    reconstruction model. With just a slight modification in the loss function, our task-adaptive approach 
    can be adapted given that a pre-trained task network is available.
    
    \item We reach state-of-the-art task performances for reconstructed scans. Both visual results and 
    metric scores are comparable with the full-dose scans for the first time in the literature in terms of 
    maintaining task-related information.

\end{itemize}

\section{Methodology}

\subsection{Overview of the Proposed Framework}

\begin{figure*}[!t]
\centerline{\includegraphics[width=\textwidth]{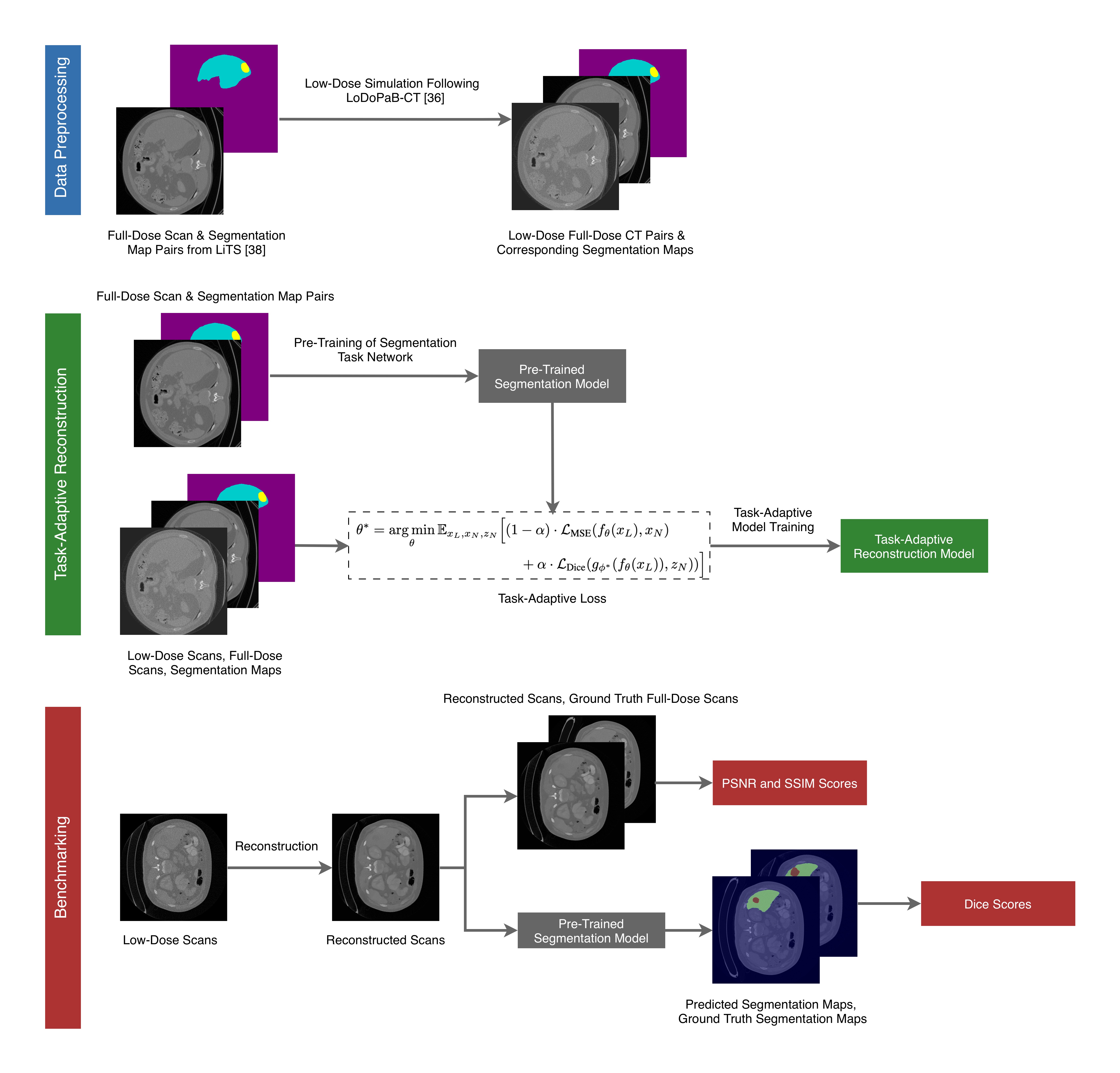}}
\caption{Overview of the proposed task-adaptive reconstruction framework with data preprocessing and benchmarking. 
Full-dose scans and ground truth task outputs are taken from a source dataset. Corresponding low-dose scans are simulated 
following Leuschner et al. \cite{lodopab-ct}. Low-dose scans are first processed by the task-adaptive reconstruction network to 
produce reconstructed images. The reconstructions are compared against the ground truth full-dose scans via the 
reconstruction loss. Simultaneously, the reconstructed images are fed into a frozen pre-trained task network to 
produce task predictions, which are compared against ground truth task outputs via the task loss. The two losses 
are combined using a weight hyperparameter, where the pre-trained task network acts as a regularizer to guide the 
training to preserve diagnostically relevant features. Reconstructed images and ground truth full-dose scans are 
compared to calculate PSNR and SSIM scores. Predicted and ground truth task outputs are compared to 
calculate the task-specific performance metric.}
\label{task-adaptive-diagram}
\end{figure*}

The proposed task-adaptive reconstruction framework follows a clear pipeline designed to preserve diagnostically relevant
information during the reconstruction process, to optimize the task performances. Low-dose CT scans serve as input to the 
reconstruction network, which produces the reconstructed images. These reconstructed images are then fed into a pre-trained task 
network that remains frozen throughout the training process. The reconstruction network is optimized using a novel loss function 
that balances reconstruction quality and task performance. The reconstruction loss measures the similarity between the reconstructed 
images and ground truth full-dose scans, while the task loss evaluates the task-specific performance by comparing the predicted task
outputs against ground truth task outputs. The weight hyperparameter $\alpha$ controls the trade-off between these two losses. 
The pre-trained task network acts as a regularizer rather than a trainable component to ensure that the reconstruction network 
learns to preserve diagnostically important features without diverging from satisfactory reconstructions. 
Fig.~\ref{task-adaptive-diagram} illustrates the complete pipeline of our proposed framework, showing the flow from input 
through reconstruction to task regularization.

\subsection{Task-Adaptive Reconstruction}

We start by showing the joint-training approach proposed by Adler et al. \cite{task-adapted-reconstruction}. 
The proposed joint loss function $\ell_{\text{joint}}$ is as follows:
\begin{equation}
    \ell_{\text{joint}}((x, d), (x', d')) := (1 - C)\ell_X(x, x') + C\ell_D(d, d')
    \label{joint-loss}
\end{equation}
for fixed $C \in [0, 1]$, where $x$ is the ground truth full-dose scan, $x'$ is the reconstructed low-dose scan, 
$d$ is the ground-truth task output, $d'$ is the predicted task output, $\ell_X$ is the reconstruction loss 
function, $\ell_D$ is the task loss function, and $C$ is the interpolation constant.

Adler et al. \cite{task-adapted-reconstruction} jointly solves the following to train 
their task adapted reconstruction model, which we call the joint-training approach:
\begin{equation}
    \resizebox{\columnwidth}{!}{%
    $(\hat{\mathcal{A}}^\dagger, \hat{\mathcal{T}}) \in \underset{\substack{\mathcal{T} \in \mathcal{M}(X,D) \\ \mathcal{A}^\dagger \in \mathcal{M}(Y,X)}}{\arg\min} \mathbb{E}_\sigma \left[ \ell_{\text{joint}} \left( (x, \tau(z)), (\mathcal{A}^\dagger(y), \mathcal{T} \circ \mathcal{A}^\dagger(y)) \right) \right]$
    }
    \label{joint-solve}
\end{equation}
where $(\hat{\mathcal{A}}^\dagger, \hat{\mathcal{T}})$ are the optimal joint estimator pair,
$\mathcal{T} \in \mathcal{M}(X,D)$ is the task estimator space, $\mathcal{A}^\dagger \in \mathcal{M}(Y,X)$ is 
the reconstruction estimator space, $\mathbb{E}_\sigma$ is the expectation over distribution $\sigma$, 
$(x, \tau(z))$ is the ground truth pair of full-dose scan and true task output, and 
$(\mathcal{A}^\dagger(y), \mathcal{T} \circ \mathcal{A}^\dagger(y))$ is predicted pair of reconstructed 
low-dose scan and predicted task output where $y$ is the observed data.

Following the mathematics of Adler et al. \cite{task-adapted-reconstruction} and incorporating the ideas 
explained in the introduction section, we propose our task-adaptive reconstruction framework. First, we 
define training of the traditional low-dose reconstruction model as the following optimization problem:
\begin{equation}
    \theta^* = \operatorname*{arg\,min}_{\theta} \mathbb{E}_{x_L, x_N} [\mathcal{L}_{\text{reconstruction}}(f_{\theta}(x_L), x_N)]
    \label{reconstruction-model}
\end{equation}
where $f_{\theta}$ is the low-dose reconstruction model for parameter space $\theta$, $x_L$ is the low-dose scan, 
$x_N$ is the ground truth full-dose scan, $\mathcal{L}_{\text{reconstruction}}$ is the reconstruction loss function, 
and $\mathbb{E}_{x_L, x_N}$ is the expectation over the dataset containing low-dose and full-dose pairs. 
The objective in \eqref{reconstruction-model} is to find the optimal set of parameters $\theta^*$ that minimizes 
the expected reconstruction loss over the entire dataset.

Then, we define pre-training of the task model as follows:
\begin{equation}
    \phi^* = \operatorname*{arg\,min}_{\phi} \mathbb{E}_{x_N, z_N} [\mathcal{L}_{\text{task}}(g_{\phi}(x_N), z_N)]
    \label{task-model}
\end{equation}
where $g_{\phi}$ is the task model for parameter space $\phi$, $z_N$ is the ground truth task output for $x_N$, 
$\mathcal{L}_{\text{task}}$ is the task loss function, and $\mathbb{E}_{x_N, z_N}$ is the expectation 
over the dataset containing full-dose scans and ground truth task outputs. The objective in \eqref{task-model} 
is to find the optimal set of parameters $\phi^*$ that minimizes the expected task loss over the entire dataset to 
pre-train the task model.

Combining \eqref{reconstruction-model} and \eqref{task-model}, we define the training of task-adaptive 
reconstruction model as the following optimization problem:
\begin{equation}
    \resizebox{\columnwidth}{!}{%
    $\begin{split}
        \theta^* = \operatorname*{arg\,min}_{\theta} \mathbb{E}_{x_L, x_N, z_N} \Big[ &(1 - \alpha) \cdot \mathcal{L}_{\text{reconstruction}}(f_{\theta}(x_L), x_N) \\
        & \quad + \alpha \cdot \mathcal{L}_{\text{task}}(g_{\phi^*}(f_{\theta}(x_L)), z_N)) \Big]
    \end{split}$
    }
    \label{task-adaptive-reconstruction-model}
\end{equation}
where $f_{\theta}$ is the task-adaptive low-dose reconstruction model for parameter space $\theta$, 
$g_{\phi^*}$ is the pre-trained task model with optimal parameters $\phi^*$, $x_L$ is the low-dose scan, 
$x_N$ is the ground truth full-dose scan, $z_N$ is the ground truth task output, 
$\mathcal{L}_{\text{reconstruction}}$ is the reconstruction loss function, 
$\mathcal{L}_{\text{task}}$ is the task loss function, $\alpha \in [0, 1]$ is the weight hyperparameter that 
controls the trade-off between the reconstruction and task losses, and $\mathbb{E}_{x_L, x_N, z_N}$ is the 
expectation over the dataset containing low-dose scans, full-dose scans, and ground truth task outputs. The 
objective in \eqref{task-adaptive-reconstruction-model} is to find the optimal set of parameters $\theta^*$ that 
minimizes the weighted sum of the reconstruction and task losses over the entire dataset. The pre-trained task 
model $g_{\phi^*}$ is frozen and acts as a regularizer to guide the reconstructions during training.

\subsection{Implementation}

The task-adaptive reconstruction model in \eqref{task-adaptive-reconstruction-model} provides the general 
form, which will be tailored for selected diagnostic task by selecting appropriate loss functions. 
In this work we focus on segmentation task to evaluate our task-adaptive reconstruction 
framework, with liver and liver tumor segmentations as the specifics. 

We use mean squared error (MSE) as our reconstruction loss following previous state-of-the-art work on LDCT reconstruction 
literature \cite{fbpconvnet}, \cite{redcnn}, 
\cite{deep-convolutional-framelet-denosing}. The MSE is defined as follows:
\begin{equation}
    \mathcal{L}_{\text{MSE}} = \frac{1}{N}\sum_{i=1}^{N} (y_i - \hat{y}_i)^2
    \label{mse-equation}
\end{equation}
where $y$ represents the ground-truth full-dose scans and $\hat{y}$ represents the reconstructed low-dose scans 
and $N$ is the total number of pixels.

We use dice loss \cite{v-net} as our segmentation loss following previous work on task-adaptive 
reconstruction literature \cite{task-adapted-reconstruction}, \cite{task-oriented}, 
\cite{empirical-evidence}. Dice score, a popular segmentation performance metric proposed for medical 
images, is used to define the dice loss \cite{v-net}. Dice score and dice loss are as follows:
\begin{equation}
    \text{Dice Score} = \frac{1}{C}\sum_{c=1}^{C} \frac{2 \sum_{i=1}^{N} p_{i,c} g_{i,c} + \epsilon}{\sum_{i=1}^{N} p_{i,c} + \sum_{i=1}^{N} g_{i,c} + \epsilon}
    \label{dice-score-equation}
\end{equation}

\begin{equation}
    \mathcal{L}_{\text{Dice}} = 1 - \text{Dice Score}
    \label{dice-loss-equation}
\end{equation}
where $C$ is the number of segmentation classes, $N$ is the number of pixels, $p_{i,c}$ is the predicted probability 
for pixel $i$ belonging to class $c$, $g_{i,c}$ is the one-hot encoded ground truth, and $\epsilon$ is a small 
smoothing factor to prevent division by zero.

Using \eqref{mse-equation} and \eqref{dice-loss-equation}, we redefine \eqref{task-adaptive-reconstruction-model} 
to obtain the following implementation that is used in the rest of our work:
\begin{equation}
    \begin{split}
        \theta^* = \operatorname*{arg\,min}_{\theta} \mathbb{E}_{x_L, x_N, z_N} \Big[ &(1 - \alpha) \cdot \mathcal{L}_{\text{MSE}}(f_{\theta}(x_L), x_N) \\
        & + \alpha \cdot \mathcal{L}_{\text{Dice}}(g_{\phi^*}(f_{\theta}(x_L)), z_N)) \Big]
    \end{split}
    \label{segmentation-adapted-reconstruction-model}
\end{equation}

\section{Experiments}

\subsection{Dataset and Preprocessing}

\begin{figure*}[!t]
\centerline{\includegraphics[width=\textwidth]{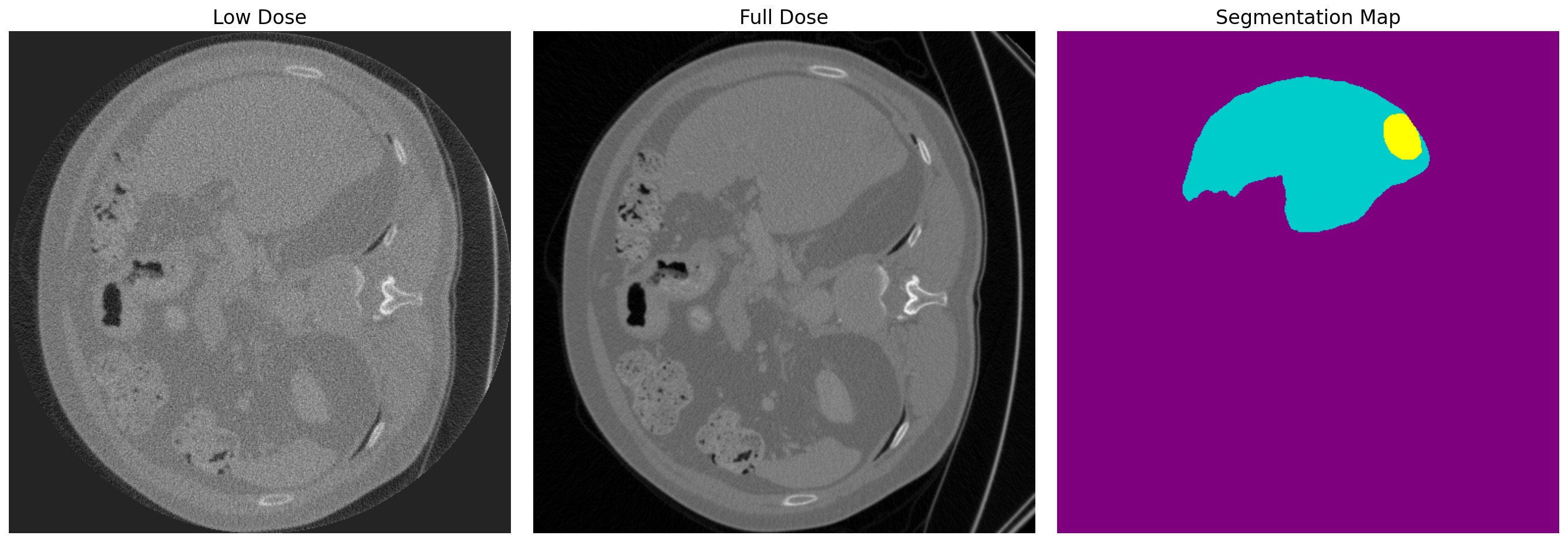}}
\caption{A sample from the preprocessed dataset showing simulated low-dose scan, original full-dose scan and 
the corresponding segmentation map. Purple area shows the background, cyan area shows the liver and yellow area 
shows the liver tumor.}
\label{dataset-sample-figure}
\end{figure*}

Due to the unavailabiliy of a dataset including low-dose and full-dose pair scans with corresponding segmentation 
maps, we created our own dataset. We first selected the publicly available Liver Tumor Segmentation Challenge (LiTS) dataset 
\cite{lits-dataset} as our source for the full-dose CT scans and corresponding segmentation maps. We then simulated 
low-dose scans from full-dose scans, following LoDoPaB-CT \cite{lodopab-ct} by Leuschner et al. to complete 
our dataset.

We could only obtain the training part of LiTS dataset which included 131 full-dose CT volumes and their 
according segmentations. In order to reduce the dataset size, we only saved the slices having non-empty segmentation 
maps while extracting the slices from the volumes. 

The preprocessing step yielded 19163 CT slice pairs, each consisting of a low-dose scan, its corresponding 
full-dose scan, and an associated non-empty segmentation map. During preprocessing, the image resolution was 
set to 512x512 pixels. We split the dataset 80\% for training and 20\% for testing purposes. 
Fig.~\ref{dataset-sample-figure} shows a sample from this 
preprocessed dataset.

\subsection{Experimental Setup}

One of our main contributions in this work, is the easy integration of our approach to existing deep 
learning-based LDCT reconstruction models. We do not have the goal of obtaining 
state-of-the-art reconstruction models. Instead, our goal is to propose a framework that is state-of-the-art 
at boosting task performances of reconstructions, and is state-of-the-art at containing task related information 
in the reconstructions.

Our codes are open-sourced and available at: \url{https://github.com/itu-biai/task_adaptive_ct}. All of our models 
apply learning rate schedulers, early stopping mechanism, configurable random seeds to ensure that we obtain the best 
performing models.

We first trained a segmentation model using ground truth full-dose scans and ground truth segmentation maps to 
obtain our pre-trained segmentation model. For our segmentation architecture, we used the original U-Net model \cite{u-net} 
with very slight modifications for increased performance, which we call "Segmentation U-Net" in the following sections.

Then, we trained a U-Net based low-dose reconstruction model using simulated low-dose scans and ground truth full-dose 
scans, to serve as our baseline model, which we call "Base U-Net". To achieve 
reconstructions with U-Net, we only added sigmoid activation function after the final layer our slightly modified U-Net, 
which we used for our segmentation U-Net pre-training. We used MSE for the loss function. 

Next, leveraging the pre-trained segmentation model, we trained our proposed task-adaptive reconstruction model 
following \eqref{segmentation-adapted-reconstruction-model} using simulated low-dose scans, ground truth full-dose 
scans and ground truth segmentation maps. We used exactly the same U-Net architecture with base U-Net. 
Only exception is that we allowed dice losses to be calculated by adding our pre-trained segmentation U-Net model 
in separate form. We trained three instances of our task-adaptive 
reconstruction model with weight hyperparameter $\alpha$ equal to 0.1, 0.5 and 0.9. Hyperparameter sweep for 
$\alpha$ was not possible due to the computational constraints, even with our significant computational resources. 
We elaborate on the computational complexity in the following sections alongside our resources and training 
approaches, and leave the details for now.

It is critical to observe that our base U-Net model is actually equivalent to our task-adaptive reconstruction model 
with weight hyperparameter $\alpha=0$. In other words, this base U-Net model is a special instance of our 
task-adaptive reconstruction which gives no weight to the segmentation task. This is an intended setup which 
enables us to directly observe the impact of our task-adaptive approach. Since both base U-Net and task-adaptive models 
share exactly the same U-Net reconstruction architecture, any performance impact on segmentation task is surely from our 
proposed task-adaptive reconstruction approach.

We also implemented the joint-training approach of Adler et al. \cite{task-adapted-reconstruction} to serve as a 
benchmark for our task-adaptive approach. To ensure fair comparison, we again used the same U-Net reconstruction 
architecture that is shared with base U-Net model and task-adaptive models. We used MSE and dice loss as the loss 
functions. We also used the same segmentation architecture that is used in segmentation U-Net model, which is 
also leveraged in the task-adaptive models in a pre-trained manner. Accordingly, we reached an architecture which 
only differs in the loss function from our task-adaptive models. We then trained three instances of the joint-training 
approach exactly following \eqref{joint-loss} and \eqref{joint-solve} with interpolation constant $C$ equal to 
0.1, 0.5 and 0.9. Just like the task-adaptive models, again hyperparameter sweep was not possible and omitted.

We also included traditional reconstruction methods for benchmarking. The first method we selected is the 
FBP (Filtered Back Projection) \cite{fbp}. It is the fundamental benchmark in LDCT reconstruction literature 
due to its computational efficiency and ease of implementation \cite{f1}. The other method we selected is BM3D 
(Block-matching and 3D Filtering) \cite{bm3d}. It is a highly powerful method that previously showed significant results, 
even competing with deep learning-based methods \cite{redcnn}. It requires a noise standard deviation 
parameter $\sigma_{psd}$ to guide the denoising process. We performed a parameter sweep on a small, randomly selected 
subset of our data, and used $\sigma_{psd}=0.075$ during our experiments accordingly.

For performance evaluation, we first decided to test our task-adaptive models' performances in PSNR 
and SSIM metrics to observe, how they compare with other models according to traditional evaluation approaches. To evaluate 
reconstruction performances, PSNR and SSIM were computed within a circular region of interest (ROI) area with a 256-pixel 
radius, centered on the 512x512 scans. This masking was necessary because the simulation methodology only generates 
data within this circular region. Consequently, the areas outside ROI are diagnostically irrelevant and appear 
uniformly gray, as illustrated in the low-dose scan in Fig.~\ref{dataset-sample-figure}. 

Finally, we first conducted low-dose reconstructions and then applied the pre-trained segmentation model. To evaluate 
the task performances, we calculated their dice scores. The dice scores are calculated using only liver and liver tumor classes 
while excluding the background class, to ensure fairness of the metric. Including the background significantly inflates the 
scores on poorly performing models while not affecting well performing models considerably.

\subsection{Resources and Computation}

We conducted all our computation for on National Center for High Performance Computing of Turkey (UHeM) 
servers under grant number 4022252025. We conducted extensive training and testing processes, utilizing a total memory 
of 27.58 TB. For the data preprocessing, we used one of UHeM's CPU servers with 2 AMD EPYC 7742 64-Core CPUs and 
distributed the job to 128 parallel processes. We conducted the training processes on UHeM's GPU servers with 4 NVIDIA A100 80GB 
VRAM GPUs and 2 AMD EPYC 7543 32-Core CPUs, running 4 distinct training setup simultaneously. We conducted the testing 
processes on the same server, but we used a hybrid approach where deep learning-based reconstructions were done on GPUs and 
traditional reconstructions were done on CPUs, all in a distributed setting.

Even with these highly powerful resources and the distributed computation setting, whole training and testing 
processes took 23 hours and 38 minutes. Accordingly, we were not able to do a hyperparameter sweep for $\alpha$ on our 
task-adaptive approach and for $C$ in joint training approach. We still tested them using values of 0.1, 0.5, 0.9 since 
their effects carried great importance for the evaluations. 

It should be noted that preprocessing step's effect on total memory utilization and total computation time is excluded. 
Accordingly, the numbers above are even slightly larger. 

\subsection{Quantitative Evaluation}

\begin{table}[t!]
    \centering
    \caption{PSNR and SSIM metric scores (mean $\pm$ std) on test set}
    \label{psnr-ssim-scores}
    \footnotesize
    \renewcommand{\arraystretch}{1.2}
    \resizebox{\columnwidth}{!}{%
        \begin{tabular}{lcc}
        \hline
        Method                       & PSNR                       & SSIM                     \\
        \hline
        Low-dose                     & 24.8596 ± 2.2471           & 0.4547 ± 0.0760          \\
        FBP                          & 28.0596 ± 2.0881           & 0.5722 ± 0.0859          \\
        BM3D                         & 34.2072 ± 3.1110           & 0.8760 ± 0.0803          \\
        Base U-Net                   & \textbf{36.0490 ± 1.3215}  & \textbf{0.9096 ± 0.0260} \\
        Joint Training $C$=0.1       & 30.8751 ± 2.1918           & 0.8750 ± 0.0250          \\
        Joint Training $C$=0.5       & 31.6969 ± 1.0815           & 0.8331 ± 0.0386          \\
        Joint Training $C$=0.9       & 21.9782 ± 2.9886           & 0.6071 ± 0.0321          \\
        Task-adaptive $\alpha$=0.1   & 32.1841 ± 1.3017           & 0.8616 ± 0.0243          \\
        Task-adaptive $\alpha$=0.5   & 30.7086 ± 1.1228           & 0.8616 ± 0.0287          \\
        Task-adaptive $\alpha$=0.9   & 31.2182 ± 1.3942           & 0.8698 ± 0.0258          \\
        \hline
        \end{tabular}
    }
\end{table}

Table~\ref{psnr-ssim-scores} shows the PSNR and SSIM metric scores for low-dose scans and all reconstruction methods. 
The low-dose scans have the lowest scores as expected. Traditional methods like FBP and BM3D improve these scores. 
BM3D results are highly effective in this sense outperforming all other methods, except Base U-Net, which achieves the
best performance in both PSNR and SSIM metrics. Still, both joint training and task-adaptive reconstruction 
approaches provide satisfactory results in these metrics, except the joint training approach with $C$ equal to 0.9. 
They show a significant improvement from the low-dose scans across both traditional evaluation metrics, PSNR and SSIM.

\begin{table}[t!]
    \centering
    \caption{Dice scores (mean $\pm$ std) of pre-trained segmentation model on test set.}
    \label{dice-scores}
    \footnotesize
    \renewcommand{\arraystretch}{1.2}
        \begin{tabular}{lcc}
        \hline
        Method                       & Dice Score                         \\
        \hline
        Low-dose                     & 0.0259 ± 0.1085                    \\
        FBP                          & 0.3065 ± 0.2935                    \\
        BM3D                         & 0.6258 ± 0.2785                    \\
        Base U-Net                   & 0.2797 ± 0.2289                    \\
        Joint Training $C$=0.1       & 0.3306 ± 0.2553                    \\
        Joint Training $C$=0.5       & 0.0129 ± 0.0493                    \\
        Joint Training $C$=0.9       & 0.0004 ± 0.0108                    \\
        Task-adaptive $\alpha$=0.1   & 0.6300 ± 0.2724                    \\
        Task-adaptive $\alpha$=0.5   & \textbf{0.7065 ± 0.2519}           \\
        Task-adaptive $\alpha$=0.9   & \textbf{0.7065 ± 0.2485}           \\
        Full-dose                    & 0.8740 ± 0.1856                    \\
        \hline
        \end{tabular}
\end{table}

Table~\ref{dice-scores} lists the Dice scores for the segmentation task using the pre-trained segmentation model. The 
score for full-dose scans represents the upper-limit performance and serves as a reference. As expected, segmentation on 
low-dose scans fails, yielding a score of almost zero. FBP improves the results from low-dose, but still falls 
significantly behind of full-dose. BM3D again provides significant results, outperforming all other methods except our 
proposed task-adaptive model. Joint training approaches with hyperparameter $C$ equal to 0.5 and 0.9 fails to conduct 
segmentations, even worse results than the low-dose scans. However, with $C$ equal to 0.1, they improve the results just 
a little further than FBP. Base U-Net method also improves the results from low-dose, but falls right behind FBP. Our 
task-adaptive models reach significant results. Our worst model with $\alpha$ equal to 0.1 improves the results from BM3D just 
a little bit. However, task-adaptive models with $\alpha$ equal to 0.5 and 0.9 reach state-of-the-art results, improving 
the segmentation task results significantly. 

\begin{figure*}[!t]
    \centerline{\includegraphics[width=\textwidth]{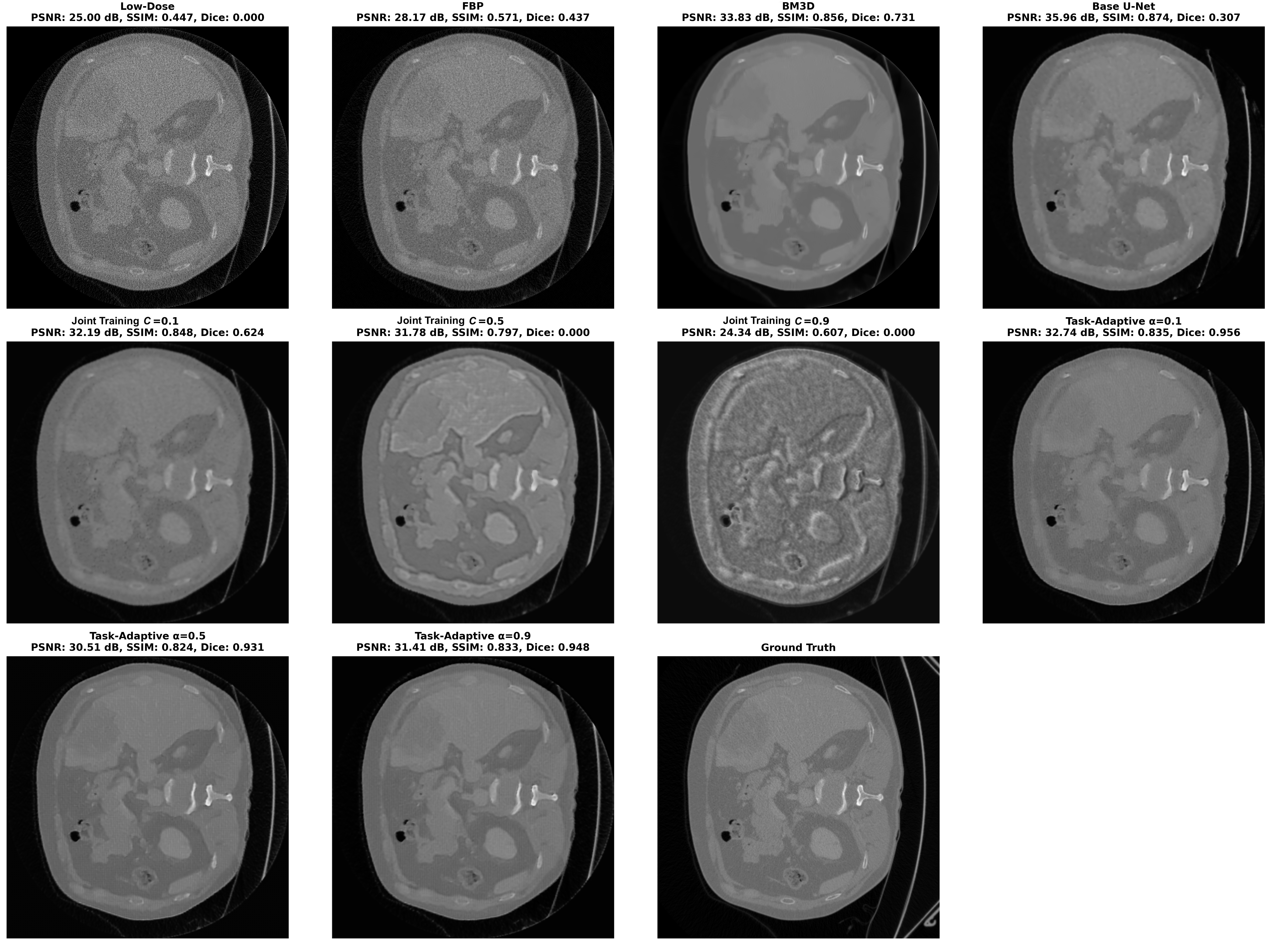}}
    \caption{A sample with low-dose scan, its corresponding reconstructions and ground truth full dose scan. 
    PSNR, SSIM and Dice scores are given above the visuals.}
    \label{reconstructions-figure}
\end{figure*}

\subsection{Qualitative Evaluation}

Fig.~\ref{reconstructions-figure} shows the reconstruction visuals alongside low-dose scan and ground truth full-dose 
scan. It can be observed that FBP improves from the low-dose scan in a very small sense, which can only be seen with a 
detailed inspection. Base U-Net's reconstruction preserves the pixel intensities of ground truth better than FBP, but the background is still 
grainy and artifacts remain. However, it is still an improvement from the low-dose scan. Joint training with $C$ equal to 
0.1 provides similar results to Base U-Net in this sense. Joint training approaches with $C$ equal to 0.5 and 0.9 
diverge from satisfactory reconstructions both losing significant diagnostic details, especially the one with 0.9. BM3D 
provides an important reconstruction that has sharp edges with clear details, yet image is oversmoothed when compared with 
the ground truth scan. Still, it contains the details better than every other method, except our task-adaptive models. All 
three instances of the task-adaptive models provide great reconstructions for this sample. However, models with 
$\alpha$ equal to 0.5 and 0.9 provide the best results, with the reconstructions being very close to ground truth full-dose 
scan.

\begin{figure*}[!t]
    \centerline{\includegraphics[width=\textwidth]{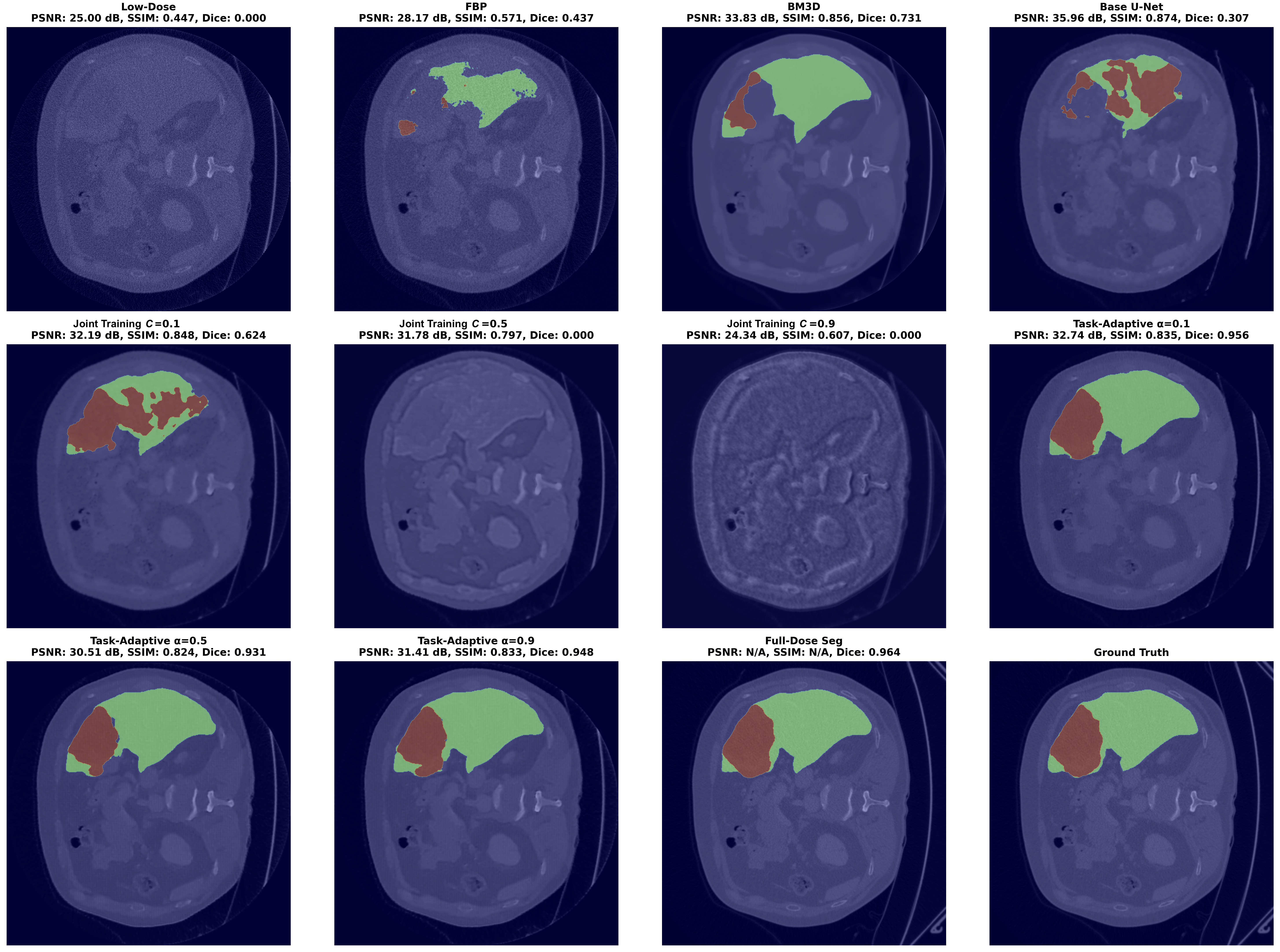}}
    \caption{Predicted segmentation maps of pre-trained segmentation model for the sample in 
    Fig.~\ref{reconstructions-figure} from the low-dose scan, its corresponding reconstructions and ground truth full dose 
    scan. Also, ground truth segmentation map is provided for reference. PSNR, SSIM and Dice scores are given above the 
    visuals.}
    \label{segmentations-figure}
\end{figure*}

Fig.~\ref{segmentations-figure} shows the predicted segmentation maps of pre-trained segmentation model for the sample in 
Fig.~\ref{reconstructions-figure}, from low-dose scan, its corresponding reconstructions and ground truth full dose 
scan, alongside with the ground truth segmentation map as reference. It is clearly seen that segmentation model failed 
to conduct any segmentation from the low-dose scan and joint training approaches with $C$ equal to 0.5 and 0.9. It also 
provided a very poor segmentation for FBP, covering just a part of the liver which is shown in green and only a 
covering a very small part of the tumor which is shown in red. The segmentation of the Base U-Net reconstruction covers 
a larger portion of the total area, but misclassifies the liver as tumor, and fails to segment the true tumor. Similarly, 
the reconstruction from the joint training approach with $C$ equal to 0.1 leads to a segmentation that captures the overall area 
well, but contains significant misclassifications between the liver and tumor. The BM3D reconstruction enables an 
highly accurate segmentation of the liver, yet it still fails to detect a big part of the tumor. Unlike other methods, all of 
our task-adaptive models produce reconstructions that yield very good segmentation results, that align well with the 
segmentation of the full-dose scan. In particular, our models with $\alpha$ equal to 0.5 and 0.9 yield 
reconstructions that produce segmentation maps, nearly indistinguishable from the ground truth. 

\section{Discussion}

The results explained in the previous section provide significant evidence that good results in PSNR and SSIM metrics 
is not sufficient for one to say that their model provides great LDCT reconstructions for clinical practice. We saw that the 
best performing model in these metrics, the Base U-Net, failed to provide satisfactory task results in segmentation task. 
Moreover, the joint training model with $C$ equals to 0.5 had a worse segmentation result than the low dose scans, which it 
completely outperformed in PSNR and SSIM metrics. Accordingly, we conclude that reconstruction models should either 
be evaluated according to new medical image quality assessment methods, or according to their task specific 
performances to observe their realistic diagnostic capabilities.

The visual results show beyond any doubt evidence that joint training approach can diverge completely from satisfactory 
results. While conducting joint training, the model simultaneously learns to reconstruct the low-dose scans and conduct 
its task. Using the segmentation task for our explanation, when interpolation hyperparameter $C$ is small, it nearly 
gives all of its focus to the reconstruction, so there is no problem about the reconstruction. Yet, its model trained 
for segmentation may perform poorly. However, when the interpolation hyperparameter $C$ is big, the model gives most of 
its focus to the segmentation results. If its jointly trained segmentation model can provide good segmentations 
from the reconstructions, the quality of the reconstructions become unimportant, as a result of its small focus in the 
loss function. Therefore, as $C$ gets bigger and bigger, reconstructions can diverge more and more from the satisfactory 
results, since its importance become even lower. The visual results of joint training approaches validate this, since the model 
with $C=0.1$ continues to provide somewhat satisfactory reconstructions while the models with $C$ equal to 0.5 and 0.9 
diverge completely from satisfactory reconstructions. Moreover, the pre-trained segmentation model that is trained with 
full-dose scan and segmentation map pairs fail to conduct any segmentation for $C$ equal to 0.5 and 0.9, while still 
conducting somewhat satisfactory segmentations for $C$ equals to 0.1, providing quantitative evidence for these ideas.

We have previously explained that the base U-Net model is actually a special instance of the task-adaptive model with 
$\alpha$ hyperparameter equal to zero. Therefore, by comparing the Base U-Net and task-adaptive models, we can directly see the effects of 
our task-adaptive models. We observe that our task-adaptive approach significantly boosts Base U-Net's task performance, 
containing more task related information visually than any other model, and providing highest dice scores among 
compared reconstruction methods. Our task adaptive approach increases the dice scores to a level that is comparable with full-dose 
segmentations, from even smaller scores than FBP and joint training with $C$ equal to 0.1. Also, our 
task-adaptive models with $\alpha$ equal to 0.5 and 0.9 are the only models that significantly outperform BM3D in the 
segmentation task. Accordingly, we state that our proposed task-adaptive approach enables LDCT reconstruction models to 
contain more task-related information than the previous approaches, resulting in better task performances. 
We conducted our experiments to validate this performance boost and investigate its reasons. In 
order to obtain state-of-the-art task performances from low-dose scans, one can train state-of-the-art 
reconstruction models with our task-adaptive approach, with also using leveraging state-of-the-art task models 
for the pre-training as well. 

\section{Conclusion}

In this work, we introduced a novel task-adaptive low-dose CT reconstruction framework that is designed to enhance 
the diagnostic performances of the resulting reconstructions. Traditional reconstruction methods, while achieving 
high scores on PSNR and SSIM metrics, often fail to preserve critical information necessary for diagnostic tasks such 
as segmentation. Our approach incorporates a pre-trained task network as a regularizer within the reconstruction model's 
loss function, controlled by a weight hyperparameter $\alpha$ that determines the trade-off between reconstruction quality and 
task performance.

Extensive experiments focusing on liver and liver tumor segmentation task, demonstrated the superiority of our 
proposed method. Our task-adaptive models significantly outperformed all other methods in terms of segmentation accuracy, 
achieving dice scores that are comparable to those obtained from full-dose scans. The framework successfully produced 
reconstructions that were both visually consistent with ground truth full-dose scans, and were also rich in task-specific information.

The main contribution of this research is the easily integratable framework that enables any existing deep learning-based 
reconstruction architecture to be optimized for a specific downstream clinical task, given that a pre-trained task network 
exists. Accordingly, this work paves the way for producing diagnostically more reliable images from low-dose CT scans. 
Future work can involve testing our framework with other diagnostic tasks, implementing it with state-of-the-art reconstruction
and task architectures, focusing on automatic tuning of the $\alpha$ hyperparameter, and exploring multi-task regularization with
multiple pre-trained task networks.

\footnotesize
\bibliographystyle{IEEEtran}
\bibliography{refs}

\end{document}